\documentclass{article}

\usepackage[final]{nips_2017}

\usepackage[utf8]{inputenc} 
\usepackage[T1]{fontenc}    
\usepackage{hyperref}       
\usepackage{url}            
\usepackage{booktabs}       
\usepackage{amsfonts}       
\usepackage{nicefrac}       
\usepackage{microtype}      

\usepackage{multirow}

\setcitestyle{numbers}
\setcitestyle{square}
\setcitestyle{comma}
\usepackage{graphicx}

\title{Detecting Behavioral Engagement of Students in the Wild Based on Contextual and Visual Data}

\author{
  Eda Okur\\
  Intel Labs\\
  Anticipatory Computing Lab \\
  Hillsboro, OR, USA \\
  \texttt{eda.okur@intel.com} \\
  \And
  Nese Alyuz \\
  Intel Labs\\
  Anticipatory Computing Lab \\
  Hillsboro, OR, USA \\
  \texttt{nese.alyuz.civitci@intel.com} \\
  \And
  Sinem Aslan \\
  Intel Labs\\
  Anticipatory Computing Lab \\
  Hillsboro, OR, USA \\
  \texttt{sinem.aslan@intel.com} \\
  \And
  Utku Genc \\
  Intel Labs\\
  Anticipatory Computing Lab \\
  Hillsboro, OR, USA \\
  \texttt{utku.genc@intel.com} \\
  \And
  Cagri Tanriover \\
  Intel Labs\\
  Anticipatory Computing Lab \\
  Hillsboro, OR, USA \\
  \texttt{cagri.tanriover@intel.com} \\
  \And
  Asli Arslan Esme \\
  Intel Labs\\
  Anticipatory Computing Lab \\
  Hillsboro, OR, USA \\
  \texttt{asli.arslan.esme@intel.com} \\
}

\begin{document}

\maketitle

\begin{abstract}
  To investigate the detection of students’ behavioral engagement (\textit{On-Task} vs. \textit{Off-Task}), we propose a two-phase approach in this study. In Phase 1, contextual logs (URLs) are utilized to assess active usage of the content platform. If there is active use, the appearance information is utilized in Phase 2 to infer behavioral engagement. Incorporating the contextual information improved the overall F1-scores from 0.77 to 0.82. Our cross-classroom and cross-platform experiments showed the proposed generic and multi-modal behavioral engagement models' applicability to a different set of students or different subject areas.
\end{abstract}

\section{Introduction}

Monitoring students’ face and upper body (appearance) as well as their interactions with the learning platform on the digital device (context) provide important cues to accurately understand different dimensions of students’ states during learning. In this study, our goal is to detect students’ behavioral engagement \cite{Fredricks-2004} (i.e., \textit{On-Task} vs. \textit{Off-Task} states) \cite{Pekrun-2012, Rodrigo-2013, Fancsali-2013} in 1:1 digital learning scenarios. Towards this end, we aim to address two research questions: (1) What level of behavioral engagement detection performance can we achieve by using a scalable multi-modal approach (i.e., camera and URL logs)? (2) How would this performance change when considering cross-subjects or cross-content platforms (Math vs. English as a Second Language (ESL))?

\section{Methodology}
\label{method}

Monitoring students’ face and upper body (appearance) as well as their interactions with the learning platform (context) provide important cues to accurately understand different dimensions of students’ states during learning. To detect behavioral engagement, we propose a two-phase system: 

\begin{enumerate}
    \item \textit{Phase 1}: Contextual data (URL logs) is processed to assess whether the student is actively using the content platform. If not (\textit{Off-Platform}), the student’s state is predicted as \textit{Off-Task}.
    \item \textit{Phase 2}: If content platform is active in learner’s device, then the appearance information is utilized to predict whether the student is \textit{On-Task} or \textit{Off-Task}.
\end{enumerate}

We trained generic appearance classifiers by employing Random Forests \cite{RF-2004} in \textit{Phase 2}. The frame-wise raw video data is used to extract face location, head position and pose, 78 facial landmark localizations, 22 facial expressions, and 7 basic facial emotions. For instance-wise feature extraction, conventional time series analysis methods were applied, such as robust statistical estimators, motion and energy measures, frequency domain features. More details regarding the appearance modality and feature extraction can be found in our previous study \cite{ICMI-2016}. Instances are sliding windows of 8-sec with 4-sec overlaps.

\section{Experimental Results}
\label{exp_res}

170 hours of multi-modal data were collected through authentic classroom pilots, from 28 9\textsuperscript{th} grade students (two different classrooms) in 22 sessions (40 minutes each), using laptops with a 3D camera. Online content platforms for two subject areas were used: (1) Math (watching videos), (2) ESL (reading articles). To obtain ground truth labels, we employed HELP \cite{ET-2017} with 3 expert labelers. We experimented with two test cases: (1) Cross-classroom, where trained models were tested on a different classroom’s data; (2) Cross-platform, where the data collected in different subject areas were utilized in training and testing, respectively. The results for these two experiments are summarized in Table~\ref{T1} and Table~\ref{T2}, respectively.

\begin{table}[!h]
  \caption{F1-scores for Cross-classroom Experiments (Set1: Classroom 1, Set2: Classroom 2, Appr: Appearance).}
  \label{T1}
  \centering
  \begin{tabular}{*5c}
    \toprule
    \textbf{Train} & \textbf{Test} & \textbf{Class} & \textbf{Appr} & \textbf{Context + Appr} \\
    \toprule
    \textbf{Set1} & \textbf{Set1} & On-Task & 0.82 & 0.82 \\
     & & Off-Task & 0.69 & 0.77 \\
     \cmidrule{3-5}
     & & Overall & 0.77 & 0.80 \\
    \midrule
    \textbf{Set1} & \textbf{Set2} & On-Task & 0.83 & 0.83 \\
     & & Off-Task & 0.63 & 0.79 \\
     \cmidrule{3-5}
     & & Overall & 0.77 & 0.82 \\
    \bottomrule
  \end{tabular}
\end{table}

\begin{table}[!h]
  \caption{F1-scores for Cross-platform Experiments (Set1: Classroom 1 with Math, Set2: Classroom 2 with Math, Set3: Classroom 1 with ESL).}
  \label{T2}
  \centering
  \begin{tabular}{*5c}
    \toprule
    \textbf{Train} & \textbf{Test} & \textbf{Class} & \textbf{Appr} & \textbf{Context + Appr} \\
    \toprule
    \textbf{Set1 + Set2} & \textbf{Set1 + Set2} & On-Task & 0.82 & 0.82 \\
    \textbf{(Math)} & \textbf{(Math)} & Off-Task & 0.67 & 0.78 \\
     \cmidrule{3-5}
     & & Overall & 0.77 & 0.80 \\
    \midrule
    \textbf{Set1 + Set2} & \textbf{Set3} & On-Task & 0.79 & - \\
    \textbf{(Math)} & \textbf{(ESL)} & Off-Task & 0.59 & - \\
     \cmidrule{3-5}
     & & Overall & 0.72 & - \\
    \bottomrule
  \end{tabular}
\end{table}

Since we have more \textit{Off-Platform} samples in Set2 than in Set1, which are predicted as \textit{Off-Task} in \textit{Phase 1}; using context improves \textit{Off-Task} scores more in Set2. We believe that the overall performance achieved is acceptable, as the expected accuracy by chance is 0.48, observed accuracy is 0.77, and Cohen’s Kappa is 0.55 for the final models. Further details of the methodology used in this study and discussions of the experimental results can be found in the full version of this paper \cite{AIED-2017}.

\section{Conclusion}

To explore scalable multi-modal approach for behavioral engagement detection, we proposed a two-phase system incorporating both visual and contextual cues. Using the context information even in the form of URL logs is rewarding for improving the overall system performance. The promising overall F1-scores show the cross-subject and cross-platform applicability of our models.

\small

\bibliography{main}

\end{document}